\documentclass{svproc}
\usepackage{url}

\usepackage{graphicx}
\usepackage{siunitx}
\usepackage{hyperref}
\usepackage[labelfont= bf]{subcaption}

\begin{document}
\mainmatter              % start of a contribution
\title{Quantitative Evaluation of Snapshot Graphs for the Analysis of Temporal Networks}
\titlerunning{Evaluation of Snapshot Graphs}  % abbreviated title (for running head)
\author{Alessandro Chiappori\inst{1} \and Rémy Cazabet\inst{2}}
\authorrunning{Alessandro Chiappori \& Rémy Cazabet} % abbreviated author list (for running head)
% \tocauthor{}
%
\institute{ENS de Lyon, 69342 Lyon Cedex 07, France\\
\email{alessandro.chiappori@ens-lyon.fr}
\and
Univ de Lyon, CNRS, Université Lyon 1, LIRIS, UMR5205,\\ F-69622 Villeurbanne, France \\ \email{remy.cazabet@gmail.com}}

\maketitle              % typeset the title of the contribution 
\begin{abstract}
One of the most common approaches to the analysis of dynamic networks is through time-window aggregation. The resulting representation is a sequence of static networks, i.e. the snapshot graph.
Despite this representation being widely used in the literature, a general framework to evaluate the soundness of snapshot graphs is still missing.
In this article, we propose two scores to quantify conflicting objectives: Stability measures how much stable the sequence of snapshots is, while Fidelity measures the loss of information compared to the original data.
We also develop a technique of targeted filtering of the links, to simplify the original temporal network. Our framework is tested on datasets of proximity and face-to-face interactions.

% We would like to encourage you to list your keywords within
% the abstract section using the \keywords{...} command.
\keywords{temporal networks, dynamic networks, time-window aggregation, snapshot graphs, choice of the window size}
\end{abstract}
\section{Introduction}
As it usually happens when studying complex systems, the analysis of temporal networks starts with the choice of the representation to work with \cite{torres2021and}.
One of the most diffused in the literature is the snapshot representation, where the original contact sequence --- a data table with rows on the form $(i, j, t)$, that we call \textit{contacts}, where $i$ and $j$ are nodes and $t$ is the associated timestep --- is converted into a sequence of static networks, that we call the \textit{snapshot graph}.
The entire time period is segmented into time windows, which can be either disjoint or partially overlapped, either with a constant or varying size \cite{leo2019non}.
For each window $[t_1, t_2]$, all the pairs of nodes with at least one contact $(i, j, t),\ t \in [t_1, t_2]$ are ``projected'' as links in the corresponding snapshot, i.e., a link $(i,j)$ is added in the snapshot representing the period $[t_1, t_2]$.
Every detail on the time ordering of the contacts inside a same snapshot is lost, although we can store as a link weight the number of instantaneous contacts (NC) that occurred over this period.

In most cases, time-window aggregation is used to create stable graphs from noisy temporal data. For instance, in a dataset of instant messages, emails, phone calls, physical proximity, etc., a static network composed of interactions occurring during a particular second, minute or hour would be composed of a very small fraction of all interactions occurring between those actors on a longer period, and would in general be completely different from the network created over the same duration during another period, even an adjacent one. On the contrary, an aggregation over several months or years would yield a very dense network that would well capture important relations, but at the cost of losing all information about the evolution of those relations over the aggregated period.
The purpose of the aggregation is thus to have windows that are large enough to ensure the dynamic stability of the contact sequence, but small enough to track an evolution in time.

It is usually taken for granted that large windows yield stable network sequences, i.e. with smooth transitions between subsequent snapshots, and thus ensure a more reliable description of the temporal network.
To test these properties quantitatively, we define two scores and perform experiments on real datasets (see Section \ref{datasets}). The results are rather different from those expectations: increasing the window size can actually bring to \textit{in}stability, so that many analysis in the literature are based on highly unstable time sequences.

To improve our scores, we propose to filter a sub-population of the links by applying a weight threshold. We show that this method can work efficiently, but only if the timescale of the filtering is longer than the timescale of the aggregation.
Finally, we conclude that the aggregation is a complex procedure that should be taken with care: all the subsequent analysis depends on the choice of the temporal network representation, despite papers in the literature often pay low attention to this very first step.

\section{Related Works}\label{biblio}
For networks of human dynamics, a common choice in the literature is to use non-overlapping windows, with fixed size corresponding to the intrinsic time scale of the dynamics --- as $\SI{24}{\hour}$ in the case of circadian periodicity.
However, the choice of the time intervals for the aggregation is often non-trivial and a same temporal network can be aggregated in different ways; this ambiguity can thus introduce biases in the analysis.
That is why researchers highlighted the influence of the choice of the window size \cite{krings2012effects,ribeiro2013quantifying} and developed methods to discover automatically the best parameters for time-window aggregation, focusing mainly on reducing biases and limiting the number of free parameters of their algorithms.

Usually, non-overlapping windows are considered, all with the same size \cite{sulo2010meaningful,leo2019non,fish2017supervised,uddin2017optimal} or with time-varying sizes \cite{soundarajan2016generating,darst2016detection,de2015structural,sun2007graphscope}.
An advantage of the snapshot representation is that single snapshots are good old static networks, so that ready-made tools can be leveraged for their analysis and visualization.
Therefore, static network measures can be used to guide the aggregation process itself \cite{krings2012effects,soundarajan2016generating,uddin2017optimal,sulo2010meaningful}.
This approach necessarily introduces biases in the resulting snapshot graph, starting from the choice of the network measure.
Supplementary biases can come from free parameters in the codes used to implement the methods.
Some of the methods make use of compression algorithms \cite{sulo2010meaningful,sun2007graphscope,de2015structural} or machine learning \cite{fish2017supervised} to perform the aggregation. We believe that these two approaches have the disadvantage of reducing the interpretability of the procedure and results.

Despite many methods having been proposed for performing automatic aggregation, a  widely-accepted method for the evaluation of the resulting snapshot graphs is still missing.
We only found three studies that outline the development of evaluation frameworks to compare different aggregating algorithms quantitatively.
In \cite{fish2017supervised}, the authors train a certain prediction task --- link prediction, attribute prediction or change-point detection --- on the snapshot graphs, splitting the entire time period into few large segments and using in turn the older segment as the training set and the newer segment as the test set.
They do not provide an absolute ranking of the methods, but different prediction tasks bring to different rankings.

An alternative framework \cite{leo2019non} focuses on the effect of the aggregation on temporal paths. Looking at the distribution of occupancy --- a property of temporal paths defined by them --- across all the links, authors suggest a threshold for the window size above which it becomes risky to trust results coming from the analysis of the snapshot graph.
Finally, \cite{cazabet2020data} uses an information theoretic approach to choose the best dynamic graph representation according to the window size, thus allowing to evaluate their relevance.

All those studies consider non-overlapping windows and warn against too large window sizes.
We believe that all those frameworks are interesting but also have drawbacks, lacking interpretability. Moreover, none of them seems to have established itself has a wide-spread reference for the following studies.
For these reasons, we found valuable to propose new properties %of our own 
for the quantitative evaluation of snapshot graphs.

\section{Proposed Evaluation Framework}
In this section we give a formal definition of our two evaluation scores, that can be applied to any temporal network.
The first is Stability, which refers to the smoothness of transitions between subsequent snapshots.
The second is Fidelity, which measures similarity between the snapshot graph and the original temporal network. 
While some forms of smoothness measures have already been proposed, to the best of our knowledge, we could not find an equivalent to Fidelity in the literature.
We chose these two measures because they propose two often conflicting objectives: we want the aggregated network to \textbf{retain as much as possible of the original information} --- Fidelity --- but in the same time, we want the aggregated graph to be \textbf{as stable as possible} --- Stability.

Stability is often an implicit objective of the aggregation: the original graph is highly unstable,  if observed at the finer scale available, and aggregation allows to obtain snapshots that are more alike from one to the next. This problem arises from the difference between the frequency of observation and the timescale of network evolution; the second timescale is usually unknown a priori.
With too short observation windows, we fail to capture the regularities of the observed phenomenon --- e.g., one snapshot per hour in an e-mail dataset will fail to capture the meaningful social network of interactions between people.
With too large windows, the network changes so much between successive snapshots that it is not possible to meaningfully relate one to the next --- e.g., 1 snapshot every 10 years in an e-mail dataset: the social relations of people have changed so much between any two successive snapshots that we do not observe an \textit{evolving} network, but rather a collection of unrelated static networks.

A parallel can be made with photography: a video recorded for a human eye requires to take approximately 24 images per seconds (24 fps). But to observe the development of a plant or the evolution of the face of someone, we should take one picture every day or month (time-lapse): for any given new picture, the subject can still be recognized easily by looking at the previous image, but there will also be some observable changes. 

Furthermore, we can note than various methods for analyzing network evolution, such as methods for community detection \cite{cazabet2020evaluating}, make the implicit assumption that the evolution of the network is smooth.

\subsection{Stability}\label{stability}
To define Stability, we first chose a measure of how much a given snapshot is akin to another one, that we call similarity. The one we mostly worked with is Jaccard coefficient \cite{krings2012effects,darst2016detection}. Given the two static networks $G_1 = (V_1, E_1)$ and $G_2 = (V_2, E_2)$ --- we use the standard graph theory formalism where $G$ is the graph, $V$ the set of vertices and $E$ the set of edges --- the Jaccard similarity between the two is:
\begin{equation}\label{eq: Jaccard}
   J(G_1, G_2) = \frac{|E_1 \cap E_2|}{|E_1 \cup E_2|} 
\end{equation}
where $|\cdot|$ computes the cardinality of a set, i.e., the number of elements in it. We easily observe that $J$ is normalized: it equals 1 if the two sets of links are identical; 0 if none of the links is in common between the two.

Other choices are possible for the similarity score, but the Jaccard coefficient is among the most diffused and we believe that it intuitively represents the idea of a smooth transition between adjacent snapshots. A recent review of network similarity scores can be found in \cite{masuda2019detecting}, including more sophisticated --- and thus less intuitive --- alternatives to the Jaccard score (based on entropy measures or graph spectral analysis, for instance).

Then, Stability $S(A)$ of the aggregated network $A$ is defined as a weighted average of snapshot-to-snapshot similarity:
\begin{equation}
    S(A) = \frac{\sum_{G_1, G_2} J(G_1, G_2) \cdot \min(|E_1|, |E_2|)}{\sum_{G_1, G_2} \min(|E_1|, |E_2|)}
\end{equation}
where the summations are over all the consecutive snapshots $G_1$ and $G_2$ in $A$. We weight similarity via the cardinality of the smallest snapshot between the two. This allows to normalize with respect to the size and also to cut out degenerate situations where one of the two snapshots has very few links, like at the beginning or at the end of a school day in the High School (HS) dataset or at the transition between weekdays and the weekend in the Copenaghen Network Study (CNS) dataset (see Section \ref{datasets}).

\subsection{Fidelity}\label{fidelity}
We quantify Fidelity by defining the Distance between the snapshot graph and the original dataset. We use a generalized graph edit distance, counting how many contacts differ between the two networks. In terms of generalized adjacency matrices --- tensors with time as the third component --- we compute the Distance $D(A)$ between the time-window aggregated network $A$ and the original network $O$ through the entry-wise $L_{1, 1, 1}$ norm:
\begin{equation}\label{eq: distance}
   D(A) = \|A - O\|_{1, 1, 1} = \sum_{i, j, t} |a_{i, j, t} - o_{i, j, t}|
\end{equation}
where entry $a_{i, j, t}$ equals 1 if the link $(i, j)$ exists in the time window that includes $t$; $o_{i, j, t}$ equals 1 if a contact was detected at the timestamp $t$. The sum is over all possible values of $i$ ,$j$ and $t$ in the original network.
$D(A)$ counts the number of disagreements --added or removed edges-- between the original and the snapshot graph. Indeed, for each contact \textit{with duration} $(i, j, [t_1, t_2])$ --- $[t_1, t_2]$ is a given window --- some of the timestamps $t \in [t_1, t_2]$ correspond to contacts in $O$, while the others correspond to false positives (FPs), i.e., to the introduction of instantaneous contacts $(i, j, t)$ that were not measured.

\section{Filtering Procedure}\label{filtering}
The idea of the filtering procedure is to cut out the links for which the real contacts cover only a small portion of the time window, i.e., those for which replacement by a contact with duration introduces too many FPs. Leaving those links out of the snapshot graph allows to improve Fidelity by reducing $D(A)$, as long as the number of false negatives (FNs) introduced --- coming from the removal of the real contacts --- is lower than the number of saved FPs.
The purpose is similar to backbone extraction for the reduction of static networks' density and noisiness \cite{coscia2017network}. In analogy with this process, we also consider temporal networks as made of a core structure of the most important links, surrounded by a cloud of less meaningful links --- the noise.
In this view, the score of Stability that we defined is expected to increase, when the noisy contacts are removed.

For the implementation, we fix a threshold value $\theta$ and only keep the links with weight greater than $\theta$, which are expected to be the most important in determining the contact structure.
We simply took the number of instantaneous contacts (NC) in the snapshot for the link weight, but other choices are possible --- as the average duration of the contacts or the number of disjoint time intervals of interaction.
Instead of taking a constant $\theta$ for all the snapshots, we fix a percentage N and we accordingly compute a corresponding $\theta$ in each window, so as to filter out the N\% of the links with the lowest weight.

The usual approach consists in filtering on each snapshot while aggregating, but we propose to consider a more general case were the two processes are independent and the filtering occurs \textit{before} the aggregation.
The idea behind this generalization is to leverage two time scales at the same time: a larger \textit{filtering window} $w_f$, ensuring that link weight distributions are not ill-defined --- i.e., to avoid limit cases where $w_f$ is only few times the measurement resolution $dt$ --- and a smaller \textit{aggregation window} $w_a$, to have a better resolution of the dynamics.
We chose to always have an integer value for the ratio $w_f / w_a$, to have a finite number of windows of the smaller size inside of the window with greater size.
We show in the next section that this configuration allows to increase stability.

\section{Results}\label{results}
\subsection{Datasets}\label{datasets}

We considered datasets of proximity and face-to-face interactions between persons, which fall into the field of human dynamics \cite{barabasi2005origin}.

The main datasets that we used for testing are the High School contact network \cite{mastrandrea2015contact} (HS) of face-to-face interactions among students and teachers, from \href{http://www.sociopatterns.org/datasets/}{SocioPatterns project}, and the Bluetooth proximity network from the Copenhagen Network Study \cite{sapiezynski2019interaction} (CNS).
The resolution of the HS dataset is $\SI{20}{\second}$; it does not correspond to the sampling frequency, but it comes from a time-window aggregation at the source.
Students and teachers from 9 classes (more then 300 persons in total) were recorded during 5 days in December 2013, using Radio Frequency Identification Devices (RFIDs).
The resolution of the CNS dataset is $\SI{5}{\minute}$. About 700 students have been involved, for a whole month.

\subsection{Stochastic Baseline}
To allow a critical interpretation of the behavior of Stability of our filtered and aggregated networks, we first present a baseline of stochastic filtering of the contacts: for every snapshot, a portion of the links with at least one contact is selected and removed from the snapshot. As a parameter, we fix the percentage value for the number of links to be removed.
We point out that stochastic sub-sampling is sometimes performed to simplify temporal networks, as in \cite{stopczynski2015temporal}.

\begin{figure}[h!t]
    \centering
    \includegraphics[width=0.6\textwidth]{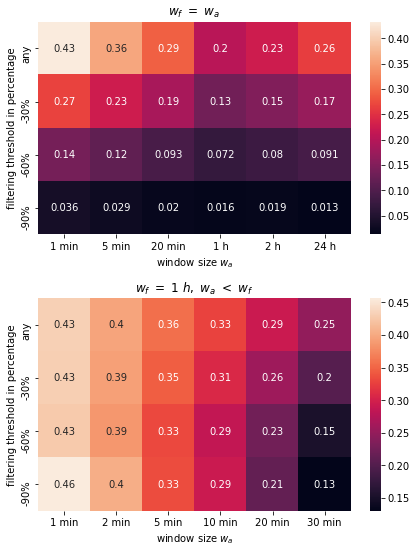}    
    \caption{\textbf{Stochastic filtering baseline.} Heatmaps of Stability score for the case $w_f = w_a$ (\textit{above}) and $w_f = \SI{1}{\hour},\ w_f > w_a$ (\textit{below}) for the HS dataset. At the value $-$N\%, for each snapshot, N\% of the links are chosen stochastically and removed}
    \label{fig:baselines}
\end{figure}

The qualitative behavior is similar between the datasets of High School (HS) and the Copenhagen Network Study (CNS) (see Section \ref{datasets}), so we only show results for the HS dataset (Figure \ref{fig:baselines}).
In the case $w_f = w_a$ stability drops while increasing the number of links removed.
Our interpretation is that, filtering stochastically, the network structure gets broken and consecutive snapshots become less similar to each other.
In the case $w_f > w_a$ --- we used $w_f = \SI{1}{\hour}$ and $w_f = \SI{2}{\hour}$ with HS and CNS, respectively --- the stability score remains constant or decreases only slightly with the filtering threshold.

We believe that this different behavior is a consequence of the particular distribution of contacts in proximity networks (see Section \ref{datasets}): very few links account for most of the contacts, i.e., the distribution of link weights is scale-free.
Hence, the stochastic filtering already illustrate the advantage of taking a larger window for the filtering than for the aggregation.

\subsection{Stability of Filtered and Aggregated Networks}

\begin{figure}[h]
    \centering
    \begin{subfigure}{0.49\textwidth}
        \includegraphics[width=\textwidth]{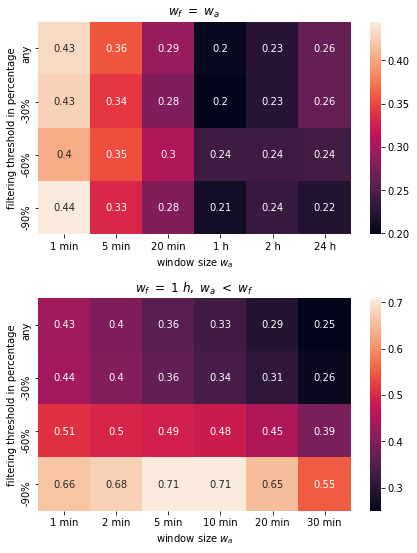}
        \caption{HS dataset}
    \end{subfigure}
    \begin{subfigure}{0.49\textwidth}
        \includegraphics[width=\textwidth]{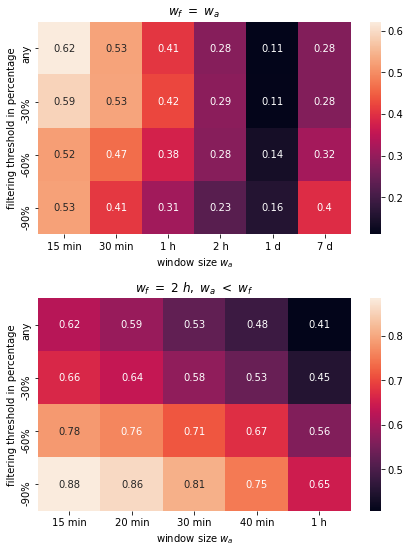}
        \caption{CNS dataset}
    \end{subfigure}
    \caption{\textbf{Stability vs filtering threshold.} Heatmaps of Stability score for the case $w_f = w_a$ (\textit{above}) and $ w_f > w_a$ (\textit{below}). At the value $-$N\%, all links with the N\% lowest weight (the NC) are removed. Links with the same weight as the first non-filtered link are also non-filtered, for consistency}
    \label{fig:heatmaps}
\end{figure}

Results for the case $w_f = w_a$ and for the case $w_f = \SI{1}{\hour},\ w_a < w_f$ with the HS dataset can be observed in Figure \ref{fig:heatmaps} (\textbf{a}).
Looking at the first row --- aggregation without filtering --- we see that the highest stability is at the smallest size, and the lowest is at intermediate sizes. The behavior is thus non trivial, with a non-monotonic dependence with respect to the window size.

A crucial aspect to notice is that absolute values in the first row are relatively low. The average Jaccard similarity is always lower than 50\% --- also at the resolution size of $dt = \SI{20}{\second}$, not displayed --- which means that more than half of the contacts change from one snapshot to the other, in average.
We are thus quite far from the idea of a motion picture with smooth transitions between frames, as we would expect from a snapshot representation.
The situation gets even worse if we consider that at the smallest time scales the snapshots are very unconventional from a network science point of view (many connected components with small size), while at higher time scales, where networks are better structured, the score drops to around 25\%. Many investigations in the literature make use of window sizes with such low values of Stability \cite{starnini2017robust,petri2014temporal,masuda2019detecting,fournet2014contact}.
Results with the CNS dataset (Figure \ref{fig:heatmaps}, \textbf{b}) are qualitatively akin.
We remark that window sizes of $\SI{24}{\hour}$ are rather common in literature \cite{krings2012effects,masuda2019detecting}, despite the low Stability.

Moving to other rows than the first one, we observe in both datasets that Stability remains more or less constant within a same column in the case $w_f = w_a$, instead of dropping with the filtering threshold as it did in the baseline.
A rather different result is found for $w_f > w_a$, where the stability score increases monotonically. High absolute values are reached in the last row, as compared to the corresponding value of aggregation without filtering reported in the first row.
Values around 70\% for HS and 80-90\% for CNS are reached when only the first decile of the most frequent links is kept (the percentage of contacts is much bigger than 10\%, around 80\% for HS and 70\% for CNS, see Section \ref{datasets}). Several observations can be made: first, at higher scales and especially for the HS dataset, Stability still remains low in absolute terms, far from the image one could have of a ``movie-like'', progressive evolution of the network.
Second, a rather heavy filtering is necessary to reach high values: how much the obtained network remains faithful to the original when only 10\% (or even less) of the links are kept? In the next section we show what Fidelity score can tell us on this issue.

\subsection{Fidelity: FPs vs FNs}\label{FPs vs FNs}
\begin{figure}[h]
    \centering
    \begin{subfigure}{0.49\textwidth}
        \includegraphics[width=1\textwidth]{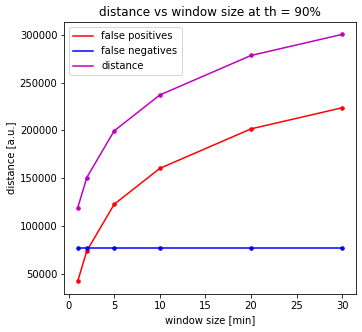}
        \caption{Fixed threshold}
    \end{subfigure}
    \begin{subfigure}{0.49\textwidth}
        \includegraphics[width=1\textwidth]{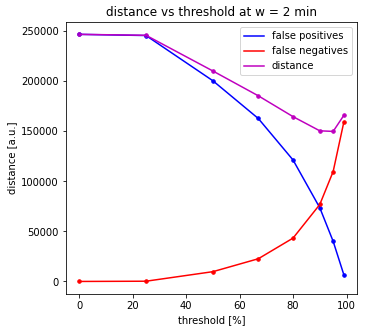}
        \caption{Fixed window size}
    \end{subfigure}
    \caption{\textbf{FPs, FNs and Distance} between original network and filtered time-window aggregated network. Data is for the configuration $w_f = \SI{1}{\hour},\ w_a < w_f$, for HS dataset}
    \label{fig: fidelity}
\end{figure}

The total Distance between the filtered-aggregated network $A$ and the original network $O$ is computed through Equation \ref{eq: distance}. Here, we divide the total Distance into false positives (FPs, contacts in $A$ but not in $O$) and false negatives (FNs, contacts in $O$ but not in $A$).
We display here two selected graphs for Distance, computed for HS dataset (Figure \ref{fig: fidelity}), while interactive 3D plots for both datasets are available online\footnote{\url{https://doi.org/10.6084/m9.figshare.16538940.v1}}.

In this configuration with fixed $w_f$, $w_a < w_f$, FNs are constant at fixed threshold --- having a same filtering window for all aggregation windows --- and FPs increase monotonically with $w_a$ (Figure \ref{fig: fidelity}, \textbf{a}).
With HS dataset, the first type of error prevails until the threshold goes beyond 90\%, then the number of FNs exceeds that of FPs, starting from the smallest sizes. This means that performing a time-window aggregation without filtering, as it is usually done in literature, plenty of FPs are introduced; but removing some of the links it is possible to reduce them to a quarter, at the cost of introducing a relatively small number of FNs.
The same results are found with the CNS datasets, but the minimumin Distance occurs at smaller values of the percentage threshold (around 67\% with $w_a$ on the order of few minutes). We think that this difference comes from the link weights distribution: for the CNS dataset it is broader, with the subset of the most frequent links being relatively less dominant.

The most interesting feature of the curves is perhaps that at high threshold the Distance between original and aggregated network presents a minimum, or a plateau (Figure \ref{fig: fidelity}, \textbf{b}). This behavior comes from the opposite monotonicity of FPs and FNs.
The behavior of the stochastic filtering baseline is much different in this high filtering limit: there is no saturation --- plateau or minimum --- but a steep decrease towards zero.

There are mainly two possible interpretations: one is that for datasets of social interaction as the ones we worked with, the system behavior can truly be well characterized by only looking at the most frequent links, neglecting most of the others. In this case, temporal networks could be simplified following our rather straightforward manner.
The second possibility is to think that the snapshot graphs obtained at high filtering actually describe only that specific subset of more frequent links, but rarer links are also crucial to characterize the network on the whole.

\section{Conclusions}
\subsection{Discussion}
We found that simple --- without filtering --- aggregation with non-overlapping windows returns small values of Stability; even smaller if the aggregation window increases. But this is precisely the most common configuration for time-window aggregation that is used in the literature, where the resulting snapshot network is next used for the analysis of temporal network's evolution.
We stress that, proceeding in this way, authors can end up working with time series that are substantially unstable, with most of the links changing from one snapshot to the next one.

Based on our observations, we propose a list of recommendations to practitioners and researchers who wish to aggregate a temporal network in snapshot graphs:
\begin{enumerate}
    \item Try several window lengths; do not choose one based on apriori, as it could lead to particularly unstable networks;
    \item Clearly state if the method/analysis presented requires or not adjacent snapshots to be stable. If it is the case, provide the Stability score;
    \item Use a threshold to filter out edges appearing a few times only in each snapshot, to limit noise influence;
    \item If the chosen aggregating window is only few time the resolution scale, try to use a larger scale for the filtering;
    \item If the contribution requires to interpret the network, also provide an analysis in terms of Fidelity of the network to the original data.
\end{enumerate}
As far as we know, we are the first to consider and prove the advantage of filtering with a larger window than the aggregation window.

\subsection{Alternatives and Future Perspectives}\label{future perspectives}
We only presented results with non-overlapping windows, but overlapping ones (sliding windows) can also be considered. They trivially increase stability, but there are at least two drawbacks: the snapshots are not independent anymore and the computational cost increases (especially if distance between two subsequent snapshots is only the resolution scale $dt$).

While we were able to validate an increase in Stability with sliding windows, reaching values only a few percentage points below 100\% even without filtering, we were not able to prove any remarkable improvement in Fidelity.
We believe that this could be due to the data structure: having a well defined timetable (lectures, breaks) for both datasets, non-overlapping windowing already give a faithful representation of the original graph.

As it is done in \cite{fish2017supervised}, some of the methods for automatic aggregation mentioned throughout Section \ref{biblio} could be chosen to test the aggregated networks that they produce, on the basis of our two scores. Then, results with our framework can be compared with alternative approaches \cite{fish2017supervised,leo2019non}.

Our scores or link weights could be modified to include time dependence of the single contacts more explicitly. For instance, the latter could decay in time, as the weights defined in \cite{holme2013epidemiologically}.

\section{Acknowledgments}
This work is supported by BITUNAM Project ANR-18-CE23-0004.

%\printbibliography
% $ biblatex auxiliary file $
% $ biblatex bbl format version 3.1 $
% Do not modify the above lines!
%
% This is an auxiliary file used by the 'biblatex' package.
% This file may safely be deleted. It will be recreated by
% biber as required.
%
%\bibliographystyle{plain}
%\bibliography{bibfile}

\end{document}